\documentclass{ws-rv975x65}
\usepackage{ws-rv-van}             
\usepackage{graphicx}
\makeindex
\begin{document}

\chapter{Scaling of von Neumann entropy at the Anderson transition}

\author[Sudip Chakravarty]{Sudip Chakravarty}

\address{Department of Physics and Astronomy,\\
University of California Los Angeles, Los Angeles, California 90024,  USA\\
sudip@physics.ucla.edu}

\begin{abstract}

Extensive body of work has shown that for the model of a non-interacting electron in a random potential there is a quantum critical point for dimensions greater than two---a metal-insulator transition. This model also plays an important role in the plateau-to-plateu transition  in the integer quantum Hall effect, which is also correctly captured by a scaling theory. Yet, in neither of these cases the ground state energy shows any non-analyticity as a function of a suitable tuning parameter, typically considered to be a hallmark of a quantum phase transition, similar to the non-analyticity of the free energy in a classical phase transition. Here we show that von Neumann entropy (entanglement entropy) is non-analytic at these phase transitions and can track the fundamental changes in the internal correlations of the ground state wave function.  In particular, it  summarizes the spatially wildly fluctuating intensities of the wave function close to the criticality of the Anderson transition. It is likely that all quantum phase transitions can be similarly described.
\end{abstract}
\body

\section{Introduction}

Ever since Anderson's paper,~\cite{Anderson:1958} ``Absence of diffusion in certain random lattices'', it has been a
theme  in condensed matter physics to unravel the  quantum phase transition between the
itinerant and the localized electronic states.~\cite{Abrahams:1979} The metal-insulator transition embodies the very basic concept of wave-particle complementarity in quantum mechanics.  
Itinerant states reflect the wave aspect, while the localized states reflect the particle aspect. In one-particle quantum mechanics without disorder, the wave and the particle descriptions are dual to each other. There is no fundamental distinction between them. Coherent superposition of waves are packets that act like spatially compact  lumps of energy and momentum, or particles. In contrast,  in a disordered medium the metallic state described by non-normalizable wave functions is separated by a quantum phase transition, the Anderson transition, when it exists,  from the insulating state with normalizable wave functions. In the insulating state particles are tied to random spatial centers. These two macroscopic states are fundamentally different and can not be analytically continued into each other.

If the Fermi energy is situated
within the localized states, the system is an insulator. It might be argued that in a real physical situation, the role of electron-electron interaction will become more and more important as the system approaches localization and the notion of Anderson localization will loose its validity. In fact, quite the opposite may sometimes  be true.  A rigorous, but a simple example of spinless fermions, was  recently studied~\cite{Shankar:1990,Schwab:2009} where interactions lead to a  broken symmetry  in the pure system, generating a gap, hence an insulator. But it was shown that for arbitrary disorder this gap is washed out, and  there are gapless localized excitations resembling an ``Anderson insulator''. In any case, Anderson transition has proven to be a powerful paradigm for metal-insulator transition.  

Because the Anderson transition is a quantum phase transition, it is natural to develop a theoretical framework that comes as close as possible to any other thermodynamical quantum phase transitions. Although there are other theoretical approaches, including powerful numerical simulations of an electron in a random potential,~\cite{MacKinnon:1981} interesting insights  can be gained by contrasting and comparing with more conventional models of phase transitions. In order to study Anderson localization I shall focus on the scaling properties of the von Neumann entropy (vNE), which is a fundamental concept in quantum mechanics and quantum information theory.  

\section{Statistical field theory of localization}
It is well known that the properties  of  a Brownian particle can be understood from a free Euclidean field theory. The free fields act as a generating function for the Brownian motion. The Green's function of interacting fields, on the other hand, reflect particles with suitable constraints.~\cite{Itzykson:1989} A particularly pretty example is that of the self avoiding random walks that can be described in terms of the correlation functions of the $O(n)$ spin model in the limit $n\to 0$, even though the partition function is exactly unity in that limit.~\cite{deGennes:1972} The lesson is that the language of statistical field theory and its scaling behavior can provide important  insights. Similarly, a replica field theory discussed elsewhere in this volume maps the Anderson problem of a single particle with disorder to  a suitable non-linear $\sigma$-model, which depends on the relevant  symmetries, with the proviso that the number of replicas $N$ has to be set to zero at the end of all calculations. It is only in the $N\to 0$ that the effect of randomness appear; as long as $N\ne 0$, the model is translationally invariant. In spite of the subtleties of the replica limit, much has been learnt  as far as the criticality of the Anderson transition is concerned by drawing analogies with the problem of critical phenomena in statistical mechanics.~\cite{Wegner:1979}

One can also reverse the chain of reasoning and learn  about the statistical mechanics of critical phenomena from the Anderson problem. As an example, let us consider the universal conductance fluctuations in a mesoscopic system. It was shown that if we consider the disorder averaged conductance by $\langle G \rangle$ and its fluctuation by $\langle (\delta G)^{2}\rangle$,~\cite{Lee:1985}  then the latter is independent of scale and is universal for dimension $D < 4$. A sample is considered to be mesoscopic if its linear dimension $L$ is larger than the mean free path  but smaller than the scale at which the phase coherence of the electrons is broken. The relative fluctuation $\langle (\delta G)^{2}\rangle/\langle G \rangle^{2}$ is proportional to $L^{4-2D}$ and is independent of scale at $D=2$.~\cite{Lee:1985} In fact,  it was shown that this result  along with many others can be obtained from a replica field theory of an extended  non-linear sigma model defined on a Grassmannian manifold.~\cite{Altshuler:1986} This raises the possibility that perhaps a similar result should also hold on a much simpler manifold, namely the coset space of $O(n)/O(n-1)$.~\cite{Chakravarty:1991} For $n=3$, this is the familiar $O(3)$ $\sigma$-model of classical $n$-vector spins of unit length $\hat{\Omega}^{2}=1$, which is a faithful description of the long wavelength behavior of the classical  Heisenberg model. What could possibly be the analog of the conductance for the Heisenberg model? It was argued that it is the spin stiffness constant, defined by the response of the system with respect to a twist in the boundary condition, which measures the rigidity of the system. By a meoscopic sample we now mean  $L$ such that it is much larger than the microscopic cutoff of the order of lattice spacing and much smaller than the correlation length  $\xi$ of the Heisenberg model. From a one-loop calculation it is easy to show that the absolute fluctuation of the spin stiffness constant $\rho_{s}$ is independent of the scale and its relative fluctuation is given by
\begin{equation}
\frac{\overline{\delta\rho_{s}^{2}}}{\overline{\rho_{s}}^{2}}\propto L^{4-2D},
\end{equation}
where the overline now represents the average with respect to the thermal fluctuations. More explicitly in the interesting case of $D=2$ we get, including the logarithmic correction,
\begin{equation}
\frac{\overline{\delta\rho_{s}^{2}}}{\overline{\rho_{s}}^{2}}=\frac{2\pi}{(n-2)[\ln (\xi/L)]^{2}}.
\end{equation}
One can find many more interesting connections between these two disparate systems, which behooves us to take a closer look at the ``thermodynamics'' of the quantum phase transition in the Anderson model, leading us to a discussion of vNE.

\section{von Neumann entropy}

A set of brief remarks seem to be appropriate to place our subsequent discussion in a more general context. In a landmark  paper on black hole entropy, Bekenstein \cite{Bekenstein:1973} demonstrated the power of the notion of information entropy. The concept also be applies to any quantum mechanical ground state. Given a unique ground state, the thermodynamic entropy is of course zero.  To distinguish various ground  states one usually studies  the  analyticity of the ground state energy as a function of a tuning parameter. In most cases, a quantum phase transition is characterized by the non-analyticity of the ground state energy.  In some cases, for example in the Anderson transition, and in the integer quantum Hall plateau transitions, the ground state energy is analytic through the transitions and does not provide any   indication of their existence. Yet we know that the wave function encodes  special correlations internal to its state. How can we quantify such correlations? In particular how do they change across these quantum  phase transitions? We shall show that in these cases the non-analyticity of vNE can be used as a fingerprint of these quantum phase transitions.~\cite{Kopp:2007}

For a pure  state $|\Psi\rangle$,  the density matrix  is $\rho=|\Psi\rangle\langle\Psi|$. Consider partitioning the system into  $A$ and $B$, where $A$ denotes the subsystem of interest and $B$ the environment whose details are of no interest. The reduced density matrix $\rho_{A}$ is constructed by tracing over the degrees of freedom of $B$, similar to integrating out the microstates corresponding to a  set of  macroscopic thermodynamic variables.  The vNE,  $S=-\mathrm{Tr (\rho_{A} \ln \rho_{A})}$, is a measure of the bipartite entanglement and therefore contains information about the quantum correlations present in the ground state. The interesting point is that the reduced density matrix is a mixture if the state $|\Psi\rangle$ is entangled, that is, it cannot be factored into $|\Psi\rangle_{A} \otimes |\Psi\rangle_{B}$. Of course, partitioning a mixed state will also lead to a mixed state; there is nothing new here. Since  $\rho_{A}$ is  a mixture, we  can perform a statistical analysis of it and obtain a non-trivial value of entropy that can summarize the essential features of an entangled state. The result follows from the Schmidt decomposition theorem:  for a bipartition of a pure  state there  exist sets of orthonormal states \{$|i_{A}\rangle$\} of $A$ and orthonormal states  \{$|i_{B}\rangle$\} of $B$ such that
\begin{equation}
|\Psi\rangle = \sum_{i} \lambda_{i}  |i_{A}\rangle \otimes  |i_{B}\rangle,
\end{equation}
where $\lambda_{i}$ are non-negative real numbers satisfying $\sum_{i}\lambda_{i}^{2}=1$. The result that a state can be fully known, yet its subsystem is in a mixed state is a remarkable consequence of entanglement. Unfortunately, there no such theorems if we partition the system into more than two parts, say $A$, $B$, and $C$. Multipartite entanglements are consequently less understood.

As we have argued above, the mapping of the Anderson localization to a problem of a  statistical field theory has been quite successful. It leaves us little doubt that the notion of criticality and scaling are correct. We might pursue this argument further and ask does this transition fit into the general framework of a quantum phase transition? If we define such a transition in terms of the non-analyticity of the ground state energy as a function of disorder, the answer to this question is no. Edwards and Thouless~\cite{Edwards:1971} have shown rigorously that the ground state energy, which depends on the average density of states, is smooth through the localization transition. We believe that the closest we can come is the non-analyticity of vNE,~\cite{Kopp:2007} which is of great current interest in regard to  quantum phase transitions.~\cite{Amico:2008}  One expects that vNE must play a role in understanding the correlations that exist on all length scales at a quantum critical point. But a state can be entangled without being critical---consider, for instance, the singlet state of two spin-$1/2$ particles. It is the special critical scaling property of entanglement that we are interested here.  Even more paradoxical, it may sound, is that Anderson localization is a single particle problem, and the conventional notion entanglement of particles does not apply. Clearly, the notion of entanglement will have to be extended, and this extension will be the theory of entanglement  defined using the site occupation number basis in the second-quantized Fock space.~\cite{Zanardi:2002}

As noted above, we shall consider two important models to illustrate our expectation. Our first example is  Anderson localization in dimension greater than two, which has been extensively studied  and is known to have a quantum critical point. At the critical point the wave function exhibits a fractal character.\cite{Wegner:1980} The second example is the plateau-to-plateau transition in the integer quantum Hall effect in which the Anderson localization plays a crucial role in establishing the very existence of the plateaus.~\cite{Laughlin:1999} We shall see that vNE is  nonanalytic at these transitions and exhibits the correct scaling behavior when compared to other approaches. vNE and its scaling behavior characterize the entanglement associated with these quantum phase transitions. Because they are determined by single-particle properties in the presence of disorder, their vNE are  different from those associated with disorder-free interacting systems.

Consider the single-particle probability $|\psi_E( r)|^2$ at energy $E$ and position $r$ for a noninteracting electronic system. In the neighborhood of a critical point governed by disorder, it fluctuates so strongly  that it  has a broad (non-Gaussian)
distribution even in the thermodynamic limit.\cite{Castellani:1986} This non-self-averaging nature of the wave function
intensity can be seen in the scaling of its moments.~\cite{Evers:2008} In particular, the moments, $P_{\ell}$, defined as the generalized inverse participation ratios, obey the finite-size scaling {\em Ansatz},
\begin{align} \label{definition of Pq}
P_{\ell}(E) \equiv \sum_{r} \overline{ \left| \psi_E(r)\right|^{2\ell
}}
\sim L^{-\tau_{\ell}} \, \mathcal{G}_{\ell}\big[(E-E_c)L^{1/
\nu}\big],
\end{align}
where $L$ is the system size and $\nu$ is the exponent characterizing the divergence of the correlation length at the critical point $E_{c}$, $\xi_E \sim
|E-E_c|^{-\nu}$. The quantity $\tau_{\ell}$ is  the multifractal spectrum, and the overline denotes the 
disorder average. $\mathcal{G}_{\ell}(x)$ is a scaling function with $\mathcal{G}_{\ell}(x\rightarrow 0) \to  1$ as  $E\to E_c$. As  $E$ deviates from the critical point, the system either tends to an ideal metallic state with
$P_{\ell}(E) \sim L^{-D(\ell -1)}$  or to a  localized state with $P_{\ell}(E)$ that is  independent
of $L$. In the  multifractal state, right at the Anderson transition, the intensity of the wave function  has local  exponents,  defined by its  sample-size dependence, which vary from point to point. A beautiful simulation of multifractality of the intensity of the wave function at the 3D Anderson transition is shown in Ref.~\citen{Vasquez:2008}.   In contrast, a single non-integer scaling exponent applicable to the entire volume corresponds to the fluctuations of a fractal. The multifractal spectrum  uniquely characterizes  the wildly complex spatial structure of the wave function. It is quite remarkable that the same multifractal spectrum determines the vNE.

\section{von Neumann Entropy in Disordered Noninteracting Electronic Systems}

We define
entanglement~\cite{Jia:2008} using the site occupation number basis in the second-quantized Fock space.~\cite{Zanardi:2002} Let us partition a
lattice of linear dimension  $L$ into two parts, $A$ and $B$. A single particle eigenstate  at
energy $E$ in  the site occupation number basis is
\begin{align}
| \psi_E \rangle &= \sum _{r \in A \cup B} \psi_E(r) \, |1
\rangle_r \bigotimes_{r' \ne r } \, |0 \rangle_{r'}
\end{align}
Here $\psi_E(r)$ is the probability amplitude  at the site $r$ and $|n\rangle_r$ is the occupation
number  at site $r$, either $0$ or $1$. We rewrite  the above sum over
lattice sites $r$ into  mutually orthogonal parts,
\begin{align}\label{decomposition of state}
| \psi_E \rangle = |1 \rangle_A \otimes |0 \rangle_{B} + |0
\rangle_A \otimes |1 \rangle_{B}
\end{align}
where
\begin{align}
|1 \rangle_A &= \sum _{r \in A} \psi_E(r) |1 \rangle_r
\bigotimes_{r'  \ne r } |0 \rangle_{r'}, \,|0 \rangle_A =
\bigotimes_{r \in A } |0 \rangle_{r}
\end{align}
similarly  for $|1 \rangle_B$ and $|0
\rangle_B$. Note that
\begin{align}
\langle 0|0 \rangle_A = \langle 0|0 \rangle_B = 1, \, \langle
1|1 \rangle_A = p_A, \, \langle 1|1 \rangle_B = p_B,
\end{align}
where
\begin{align}
 p_{A}= \sum_{r \in A}  |\psi_E(r)|^2,
\end{align}
and similarly for $p_B$ with $p_A + p_B =1$.

The reduced density matrix $\rho_A$ is obtained from  $\rho = | \psi_E
\rangle \langle \psi_E |$, after tracing out the
Hilbert space over $B$, is
\begin{align}
\rho_A  & = |1 \rangle_A \langle 1| + (1-p_{A}) |0 \rangle_A \langle 0|.
\end{align}
The corresponding vNE is given by
\begin{align}\label{bipartite entanglement}
S_A = - p_A \ln p_A -(1- p_A) \ln (1-p_{A}).
\end{align}
Here,  manifestly $S_A = S_B$, and either of them is bounded between $0$ and $\ln 2$
for any eigenstate.  Despite the  use of  a second-quantized
language, we are considering  a single particle state rather than a many body correlated state. The
entanglement entropy can not grow arbitrarily large as  the size of $A$ increases, unlike the entanglement entropy in interacting quantum systems where it can be arbitrarily large close to the critical point. 

If the system size becomes very large in comparison to the size of the subsystem $A$,
we can restrict $A$  to be a single lattice site and study scaling with respect to $L$.  Thus, we consider the single site vNE~\cite{Kopp:2007}
\begin{align}\label{single site entropy sum}
S(E)  &= - \sum_{r \in L^d} \Bigl\{ |\psi_E(r)|^2   \ln |\psi_E(r)|^2 \nonumber \\ & \quad + \left[1- |\psi_E(r)|^2
\right] \ln \left[ 1- |\psi_E(r)|^2 \right]\Bigr\}.
\end{align}

To study the leading critical behavior, the second term in the curly brackets in the right-hand side of Eq. \eqref{single site entropy sum} can be ignored since
$\left| \psi_{E}(r)\right|^2  \ll 1$ for all $r$ for states close to the critical energy. 
The disorder averaged entropy $\overline S$ can be expressed in terms of the  multifractal scaling in Eq. \eqref{definition
of Pq}, giving
\begin{align}\label{EE summed over all sites}
\overline{S}(E)\approx -\frac{dP_{\ell}}{d\ell}\bigg|_{\ell=1} \approx
\frac{d\tau_\ell}{d\ell}\bigg|_{\ell =1} \ln L - \frac{\partial
\mathcal{G}_\ell}{\partial \ell} \bigg|_{\ell=1}.
\end{align}
Although we do not know the anaytical form of the scaling function $\mathcal{G}_\ell$, its approximate $L$ dependence
can be obtained  in various limiting cases. Exactly at criticality, $\mathcal{G}_\ell \equiv 1$ for all values of
$\ell$ and
\begin{align}\label{EE critical scaling}
\overline{S}(E) \sim \alpha_1 \ln L,
\end{align}
where  the constant $\alpha_1 ={d\tau_{\ell}/d\ell}|_{\ell =1}$.
The leading scaling
behaviors of $\overline{S}(E)$ in both the metallic and the localized states can now be obtained, following the discussion below Eq. \eqref{definition of Pq}. The results are
\begin{align}\label{leading}
\overline{S}_{metal}(E) \sim D \ln L, \;\;\;\; \overline{S}_{loc}(E)  \sim \alpha_1 \ln
\xi_E.
\end{align}
We see that in general $\overline{S}(E)$ is of the  form
\begin{align}\label{approximate form for EE of single energy state}
\overline{S}(E) \sim \mathcal{Q}[(E-E_C)L^{1/\nu}] \ln L,
\end{align}
where the coefficient function $\mathcal{Q}(x)$  is
$D$ in the metallic state, decreases to $\alpha_1$ at criticality and then
goes to zero for the localized state. We now turn to numerical simulations to see the extent to which this scaling
behavior is satisfied.

\section{von Neumann entropy  in the three dimensional Anderson Model\label{sec:Amodel}}

Let us consider the disordered
Anderson model on a 3D cubic lattice.~\cite{Jia:2008} The Hamiltonian is
\begin{equation}\label{Hamiltonian_Anderson}
    H=\sum_i V_i c_i^\dag c_i-t\sum_{\langle i,j\rangle}(c_i^\dag
    c_j+H.c.),
\end{equation}
where $c_i^{\dag}$($c_i$) is the creation (annihilation) operator for an electron at site $i$ and the $\langle i,j \rangle$ indicates that the second sum is over nearest neighbors. The $V_i$ are random variables uniformly distributed in the range
$[-W/2,W/2]$.  In what follows, we set $t=1$. Of course, the model has been extensively studied. Below a critical disorder strength $W_c$, there is a region of extended states at the band center.\cite{MacKinnon:1981} The recent values of the critical disorder strength $W_{c}$ and the
localization length exponent are $W_{c}=16.3$ and $\nu=1.57\pm 0.03$. \cite{Slevin:2001} 

To obtain the energy-averaged entropy, we average Eq.
\eqref{single site entropy sum} over the  entire  band of
energy eigenvalues. From this we construct the vNE,
\begin{align}
\overline{S}(w,L) = \frac{1}{\cal N}\sum_{E} \overline{S}(E,w,L),
\end{align}
where $\cal N$ counts the total number of states in the band. Near $w=0$, we can show, using  Eqs. \eqref{approximate form for EE of single
energy state} and \eqref{generalscaling2}, that
\begin{align}\label{generalscaling1}
\overline{S}(w,L) \sim C +L^{-1/\nu}f_{\pm}\big(wL^{1/\nu}\big)\ln L,
\end{align}
where $C$ is a constant independent of $L$
 and $f_\pm(x)$
are two universal functions corresponding to the regimes
$w>0$ and $w<0$.
\begin{figure}
    \centering
    \includegraphics[width=\columnwidth]{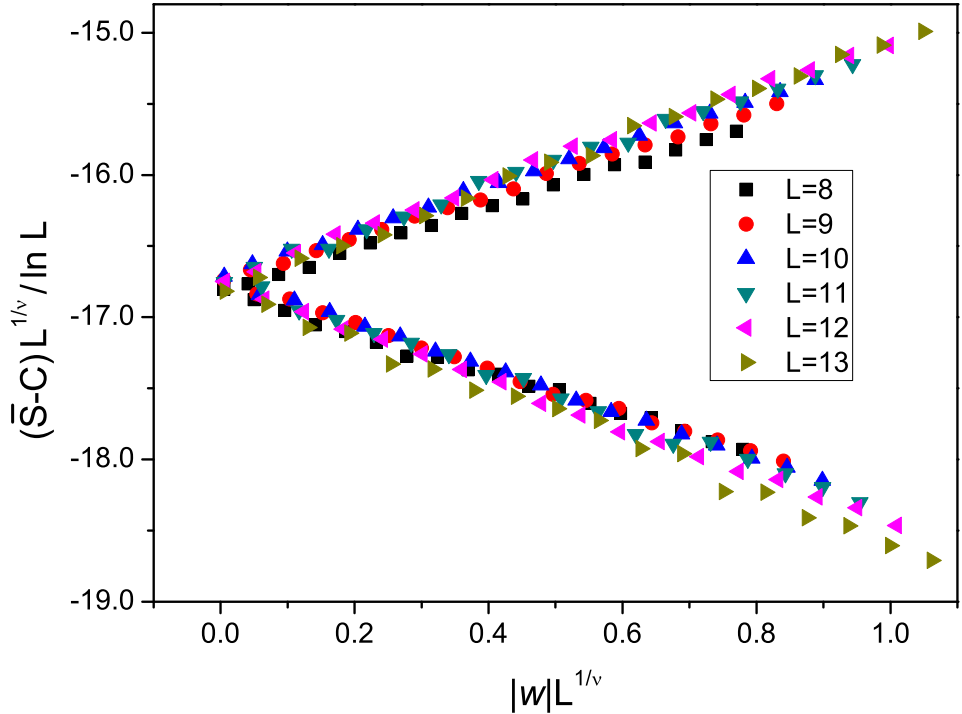}\\
    \caption{Scaling curve in the 3D Anderson model. With the
    choice of $\nu=1.57$ and $C=12.96$, all data collapse to a
    universal functions $f_\pm(x)$. The two branches correspond
    to $w<0$ and $w>0$.} \label{Scaling_Anderson1}
\end{figure}
We numerically diagonalize Eq. (\ref{Hamiltonian_Anderson}) for systems of sizes  $L\times L\times L$ with
periodic boundary conditions.  The maximum system size was $L=13$, and the results were averaged over 20 disorder
realizations. The scaling form of $\overline{S}(w,L)$ is given by Eq. \eqref{generalscaling1}.
Figure~\ref{Scaling_Anderson1} shows~\cite{Jia:2008}  the results of the data collapse with a choice of $\nu=1.57$, and the nonuniversal
constant $C=12.96$ is determined by a powerful algorithm described in the Appendix C of Ref.~\citen{Goswami:2007}. The 
data collapse is reasonable and  is consistent with  the nonanalyticity of vNE and the multifractal analysis. Clearly, it would be useful to improve the numerics by increasing both the system sizes and the number of disorder realizations to attain a better data collapse.

\begin{figure}
\centering
  \includegraphics[width=\columnwidth]{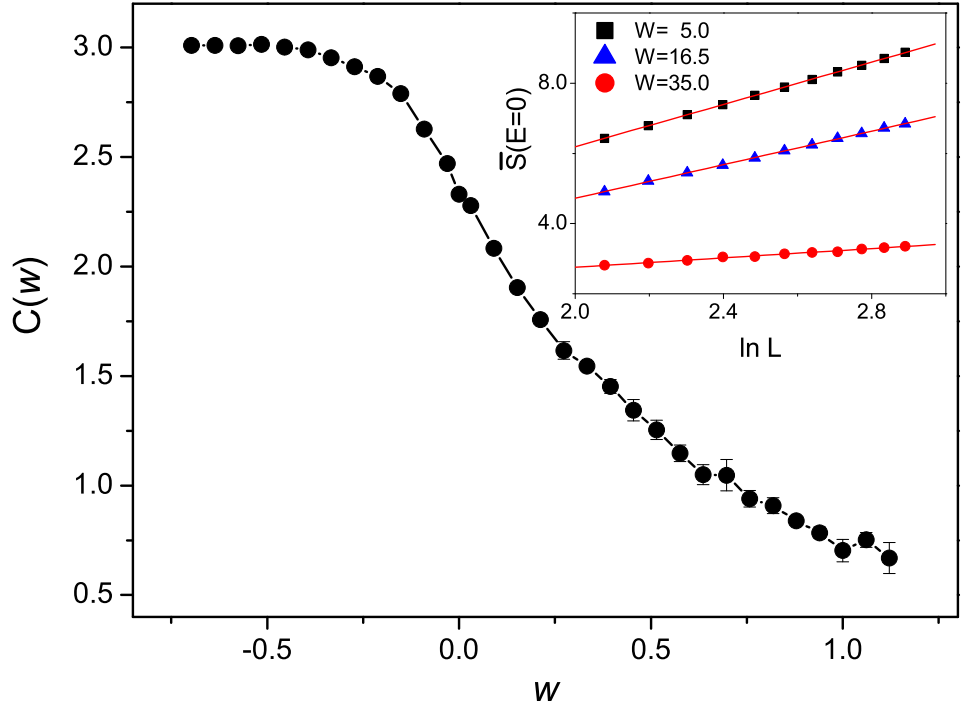}\\
  \caption{The quantity ${\cal C}$ in
  Eq.~(\ref{generalscaling2}). The  system sizes
 are too small to observe the weak $L$ dependence. Inset:
$\overline{S}(E=0,W,L)$ as a function of $\ln L$ for three different $W$.}
  \label{scaling_Anderson2}
\end{figure}

We can also study vNE at the  band center $E=0$ by sweeping $W$
across the critical value $W_c$. In this
case, the states at $E=0$ will evolve continuously from
metallic to critical and then to localized states.  The entanglement  entropy will be given similarly by another scaling function
\begin{align}\label{generalscaling2}
\overline{S}(E=0,w,L) \sim \mathcal{C}(wL^{1/\nu})\ln L,
\end{align}
where  $w=(W-W_c)/W_c$ is the reduced disorder
strength and $\mathcal{C}(x)$ is a scaling function, which as remarked earlier, $ \to D$ as $w
\to -1$ and $ \to 0$ as $w \to \infty$, and
$\mathcal{C} = \alpha_1$ when $w=0$.
For this purpose we use the transfer matrix method \cite{Kramer:1996} to study  the energy resolved $\overline{S}(E,w,L)$ by
considering a quasi-one-dimensional  system with a size of $(mL)\times L\times L$, $m\gg 1$;    $L$ up
to $18$, and $m=2000$ were found to be reasonable. To compute vNE, we divide the system into $m$ cubes
labeled by $I=1,2,\ldots,m$, each containing $L^3$ sites. The wave function within each cube is normalized and
the vNE, $\overline{S^I}(E,W,L)$ in the $I^{\text{th}}$ cube is computed. Finally $\overline{S}(E,W,L)$ was obtained by
averaging over all cubes.
 The validity of the scaling form in
Eq.~(\ref{generalscaling2}) is seen in Fig.~\ref{scaling_Anderson2}.~\cite{Jia:2008} In particular, the function $\mathcal C(x)$ shows
the expected behavior, approaching $D=3$ as $w \to -1$, and tending to $0$ as $w\to \infty$ .

\section{von Neumann entropy in the integer quantum Hall system}

For  the integer quantum Hall system, we use a basis defined by the states  $|n,k\rangle$, where  $n$ is the  Landau
level index and  $k$ is the wave vector in the $y$-direction. The Hamiltonian can be
expressed~\cite{Huckestein:1995} in terms of the matrix elements in this basis as 
\begin{equation}
      H=\sum_{n,k}|n,k\rangle\langle
    n,k|\left(n+\frac{1}{2}\right)\hbar\omega_c
    +\sum_{n,k}\sum_{n',k'}|n,k\rangle\langle n,k|V|n',k'\rangle\langle n',k'|
\end{equation}
where $\omega_c=eB/mc$ is the cyclotron
frequency, and $B$ is the magnetic field. $V(\mathbf{r})$ is the
disorder potential. If we focus on the lowest Landau level,
$n=0$,  and assume that the distribution of disorder is $\delta$-correlated with zero mean, that is, $\overline{V(\mathbf{r})}=0$ and 
$\overline{V(\mathbf{r})V(\mathbf{r'})}=V_0^2\delta(\mathbf{r}-\mathbf{r'})$,
the matrix elements, $\langle 0,k|V|0,k'\rangle$, are~\cite{Huckestein:1995}
\begin{equation}
    \langle 0,k|V|0,k'\rangle=\frac{V_0}{\sqrt{\pi
    L_y}}\exp\left[-\frac{1}{4}l_B^2(k-k')^2\right]
    \int\mathrm{d}\chi e^{-\chi^2}u_0(l_B\chi+\frac{k+k'}{2}l_B^2,k'-k)
    \label{matrixelement}
\end{equation}
where $l_B=(\hbar c/eB)^{1/2}$ is the magnetic length, and $u_0(x,k)$
is the Fourier transform of  $V(x,y)$ along the $y$ direction,
\begin{equation}
u_{0}(x,k)=\frac{1}{\sqrt{L_{y}}}\int dy V(x,y) e^{iky}.
\end{equation}

We choose a  two-dimensional square with a linear dimension $L=\sqrt{2\pi} Ml_B$, where $M$ is an integer. We impose periodic boundary conditions in both directions and discretize by  a mesh of size $\sqrt{\pi} l_B/\sqrt{2}M$. The Hamiltonian matrix  is diagonalized and the
eigenstates $|\phi_a\rangle=\sum_k \alpha_{k,a}|0,k\rangle, \; a=1 \ldots M^2$, are obtained along with the corresponding eigenvalues
$E_a$. The zero of the energy is at the center of the lowest Landau band \cite{Ando:1974} and the unit of energy is $\Gamma=2V_0/\sqrt{2\pi}l_B$. For each eigenstate the wave function in real space is constructed:
\begin{equation}
    \phi_a(x,y)=\langle x,y|\phi_a\rangle=\sum_k
    \alpha_{k,a}\phi_{0,k}(x,y), \label{wave function_IQHE}
\end{equation}
where $\phi_{0,k}(x,y)$ is the wave function with  quantum number $k$ in the lowest Landau level.
The dimension of the Hamiltonian matrix increases as $N_k\sim M^2$, making it difficult to diagonalize fully. We circumvent this difficulty~\cite{Jia:2008}   by computing only those states $|\phi_a\rangle$ whose energies lie within a window $\Delta$ around a fixed value $E$ thus:
$E_a\in[E-\Delta/2,E+\Delta/2]$. We take $\Delta$ to be sufficiently small ($0.01$), but still large
enough such it spans  large number of states in the interval $\Delta$ (at least 100 eigenstates).

\begin{figure}
 \centering
  \includegraphics[width=\columnwidth]{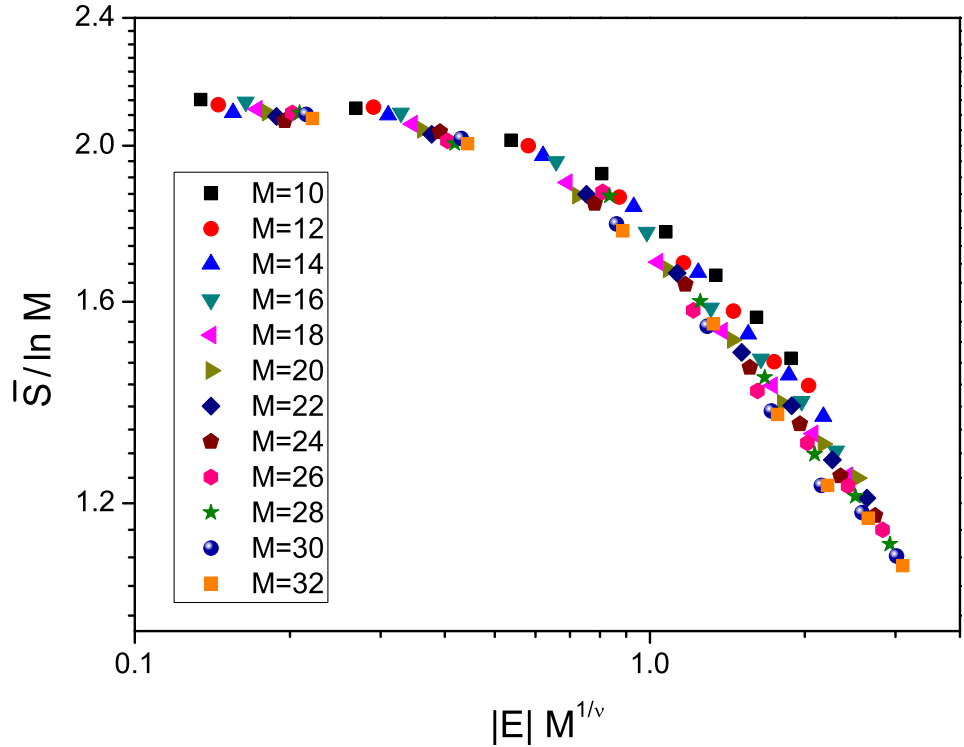}
  \caption{Scaling of the von Neumann entropy  $\overline{S}(E)$ for the integer quantum Hall effect. $M$ instead of $L$ is used in the data
  collapse with the accepted value of $\nu\approx 2.33$; see Ref.~\citen{Huckestein:1995}.}
  \label{scaling_IQHE}
\end{figure}

We now uniformly break up the $L\times L$ square into smaller squares $\mathcal{A}_i$ of size $l\times l$, where
$l=l_B\sqrt{\pi/2}$, independent of the system size $L$. The $\mathcal{A}_i$ do not overlap. For each of the states, we compute the coarse grained quantity
$\int_{(x,y)\in\mathcal{A}_i}|\psi_a(x,y)|^2\mathrm{d}x\mathrm{d}y$. The vNE for a given eigenstate
is calculated by the same procedure described above for the Anderson localization.  The vNE $\overline{S}(E, L)$ is obtained at energy $E$ by averaging over states in the interval
$\Delta$. $\overline{S}(E,L)$ has a scaling form
given by Eq. \eqref{approximate form for EE of single energy state} with $E_c = 0$; it is
$\overline{S}(E,L)=\mathcal{K}(|E|L^{1/\nu})\ln L$. Figure ~\ref{scaling_IQHE} shows reasonably good agreement with the numerical simulations.~\cite{Jia:2008} The exponent $\nu$ is consistent with that obtained by other approaches.~\cite{Huckestein:1995} Thus, the criticality of the vNE  at the center of the Landau band  is demonstrated. There is only one branch of the scaling curve because all states are localized except at the center of the band. Again, more extensive numerical calculations are necessary to obtain more definitive results.

\section{A brief note on the single-site von Neumann entropy}

The scaling of single-site vNE can lead to some misunderstanding in regard to universality. This can be illustrated by considering   Ising chain in a transverse field for which the Hamiltonian is 
\begin{equation}
{\cal H} = - J\lambda\sum_i S_i^z S_{i+1}^z-J\sum_i S_i^x
\end{equation}
where $S_{x}$, $S_{y}$ and $S_{z}$ are spin-1/2 matrices. The sum is over all sites $N\to \infty$.
It is well known that the second derivative of the ground state energy~\cite{Pfeuty:1970}
\begin{equation}
\frac{E_{0}}{N}=-\frac{2J}{\pi}(1+\lambda)\mathbb{E}\left(\frac{2\sqrt{\lambda}}{1+\lambda}\right)
\end{equation}
has a logarithmic singularity at $\lambda=1$, signifying a quantum critical point in the conventional sense of the non-analyticity of the ground state energy. This non-analyticity  is symmetric as $\lambda\to 1\pm$ . Here, $\mathbb{E}$ is the complete Elliptic integral of the second kind. It is also simple to calculate the single site vNE. A  given site constitutes part $A$ of the system, and part $B$ is the rest of the Ising chain of $N-1$ sites. The reduced density matrix $\rho_{A}$ is~\cite{Osborne:2002}
\begin{equation}
\rho_A=\frac{1}{2}\left(\begin{array}{cc}1+\langle \sigma^{z}_{i}\rangle &  \langle \sigma^{x}_{i}\rangle \\ \langle \sigma^{x}_{i}\rangle& 1-\langle \sigma^{z}_{i}\rangle\end{array}\right),
\end{equation}
where the exact known results are~\cite{Pfeuty:1970} 
\begin{eqnarray}
\langle\sigma^z_i\rangle &=& (1-1/\lambda^2)^{1/8}, \; \lambda > 1; 0, \; \textrm{otherwise},\\
\langle \sigma^x_i\rangle &=&\frac{1-\lambda}{\pi}{\mathbb K}\left( \frac{2\sqrt{\lambda}}{1+\lambda}\right)+\frac{1+\lambda}{\pi}{\mathbb E}\left( \frac{2\sqrt{\lambda}}{1+\lambda}\right),
\end{eqnarray}
where $\mathbb K$ is the complete elliptic integral of the first kind. The vNE, $S$, can now be easily computed from the $2\times 2$ reduced density matrix $\rho_{A}$.
The singularities approaching the critical point are
\begin{eqnarray}
\lim_{\lambda \to 1-}\frac{\partial S}{\partial \lambda}&=&-\frac{1}{2\pi}\ln\left(\frac{\pi+2}{\pi-2}\right)\ln |\lambda -1|, \\
\lim_{\lambda \to 1+}\frac{\partial S}{\partial \lambda}&=&-\frac{\pi}{2^{19/4}}\ln\left(\frac{\pi+2}{\pi-2}\right)(\lambda-1)^{-3/4}
\end{eqnarray}
The exponents differ as to how  we approach the critical point. Nonetheless,  the exponents are pure numbers independent of the coupling constant, as are the amplitudes. This then is  a perfectly legitimate case of universality. The reason for the  asymmetry at the critical point is clear: the magnetization (a local order parameter) vanishes for $\lambda \le 1$, while it is non-zero for $\lambda >1$. For the case of Anderson localization and the integer quantum Hall effect, there are no such local order parameters that vanish at the transition. If we regard the average density of states as an order parameter, it is smooth through both the Anderson transition and the plateau-to-plateau transition for the integer quantum Hall effect. Thus, the single site vNE has a scaling function that is symmetric around the transition as deduced from the multifractal scaling. The moral is that the single site entropy is an important and useful quantity to compute.

\section{Epilogue}

Entropy measures uncertainty in a physical system. It is therefore not surprising that it is a central concept in quantum information theory. That it may turn be an essential concept at a quantum critical point can also be anticipated. At a critical point a system cannot decide in which phase it should be. At the Anderson transition the wave function is a highly complex multifractal, and it is not surprising that vNE exhibits non-analyticity in the infinite volume limit, even though the ground state energy in which the complexity of the wave function is averaged over is smooth through it.  The non-analyticity of vNE is perfectly consistent with  other measures of entanglement, for example the linear entropy, \cite{Zurek:1993} which is $S_{L}=1-\mathrm{Tr \rho_{A}^{2}}$. The inverse participation ratio $P^{(2)}=\frac{1}{\cal N}\sum_{r,E}|\psi_{r}(E)|^{4}=1-S_{L}/2$, where $\cal N$ is the total number of states in the band.  In the extreme localized case, only one site participates and $S_{L}=0$. In the opposite limit  $S_{L}=2-2/N\to 2$, when $N\to \infty$. The participation ratio, hence $S_{L}$, exhibits scaling at the metal-insulator transition: $P^{(2)}=L^{-x} g_{\pm}(L^{1/\nu}w)$, where $g_{\pm}$ is another universal function. This scaling was also verified with $x\approx 1.4$, $\nu \approx 1.35$, and $W_{c}\approx 16.5$ for the $3D$ Anderson transition.~\cite{Kopp:2007}

A key question now is what happens when we have both disorder and interaction. In recent years there has been much progress in one-dimensional systems, especially from the perspective of vNE in the ground state.~\cite{Refael:2009} The quantum criticalities   of these disordered interacting systems belong to universality classes different from their counterparts with interaction but no disorder and are generally described by infinite disorder fixed points. Little is understood for similar  higher dimensional systems. The principal difficulty in constructing a universal theory when both interactions and disorder are present is qualitatively clear. When interactions are strong and disorder is absent, the ground state can break many symmetries and organize itself into a variety of phases. Introducing disorder may affect the stability of these many body correlated phases in different ways.~\cite{Chakravarty:1999,Voelker:2001} Although the symmetry of the order parameters can guide us, the strongly correlated nature of these phases makes theories difficult to control. As mentioned above, in a simple case of spinless fermion,  we were able to provide some rigorous answers: no matter how strong the interaction is there appears to be gapless excited states and the broken symmetry is broken. In the opposite limit, we have to examine  how weak interaction affects the Anderson problem. Here there has been progress in recent years; see the contribution by Finkelstein in the present volume.

\section*{Acknowledgements}
I would like to thank my collaborators A. Kopp, X. Jia, A. Subramanium, D. J. Schwab, and I. Gruzberg. This work was supported by a grant from the National Science Foundation, DMR-0705092.


\begin{thebibliography}{33}
\providecommand{\natexlab}[1]{#1}
\providecommand{\url}[1]{\texttt{#1}}
\expandafter\ifx\csname urlstyle\endcsname\relax
  \providecommand{\doi}[1]{doi: #1}\else
  \providecommand{\doi}{doi: \begingroup \urlstyle{rm}\Url}\fi

\bibitem{Anderson:1958}
P.~W. Anderson, Absence of diffusion in certain random lattices, \emph{Phys.
  Rev.} {\bf 109}, \penalty0 1492,  (1958).

\bibitem{Abrahams:1979}
E.~Abrahams, P.~W. Anderson, D.~C. Licciardello, and T.~V. Ramakrishnan,
  Scaling theory of localization: Absence of quantum diffusion in two
  dimensions, \emph{Phys. Rev. Lett.} {\bf 42}, \penalty0 673--676,  (1979).

\bibitem{Shankar:1990}
R.~Shankar, Solvable model of a metal-insulator transition, \emph{Int. J. Mod.
  Phys. B}. {\bf 4}, \penalty0 2371--2394,  (1990).

\bibitem{Schwab:2009}
D.~J. Schwab and S.~Chakravarty, Glassy states in fermionic systems with strong
  disorder and interactions, \emph{Phys. Rev. B}. {\bf 79}, \penalty0 125102,
  (2009).

\bibitem{MacKinnon:1981}
A.~MacKinnon and B.~Kramer, One-parameter scaling of localization length and
  conductance in disordered systems, \emph{Phys. Rev. Lett.} {\bf 47},
  \penalty0 1546--1549,  (1981).

\bibitem{Itzykson:1989}
C.~Itzykson and J.-M. Drouffe, \emph{Statistical Field Theory}. vol.~1,
  (Cambridge University Press, Cambridge, 1989).

\bibitem{deGennes:1972}
P.~G. de~Gennes, Exponents for the excluded volume problem as derived by the
  {W}ilson method, \emph{Physics Letters A}. {\bf 38}, \penalty0 339--340,
  (1972).

\bibitem{Wegner:1979}
F.~Wegner, Mobility edge problem - continuous symmetry and a conjecture,
  \emph{Z. Phys. B}. {\bf 35}, \penalty0 207--210,  (1979).

\bibitem{Lee:1985}
P.~A. Lee and A.~D. Stone, Universal conductance fluctuations in metals,
  \emph{Phys. Rev. Lett.} {\bf 55}, \penalty0 1622,  (1985).

\bibitem{Altshuler:1986}
B.~L. Altshuler, V.~E. Kravtsov, and I.~V. Lerner, Statistics of mesoscopic
  fluctuations and instability of one-parameter scaling, \emph{Zh. Eksp. Teor.
  Fiz.} {\bf 91}, \penalty0 2276--2302,  (1986).
\newblock Sov. Phys. JETP {\bf 64}, 1352 (1986).

\bibitem{Chakravarty:1991}
S.~Chakravarty, Scale-independent fluctuations of spin stiffness in the
  {H}eisenberg model and its relationship to universal conductance
  fluctuations, \emph{Phys. Rev. Lett.} {\bf 66}, \penalty0 481,  (1991).

\bibitem{Bekenstein:1973}
J.~D. Bekenstein, Black holes and entropy, \emph{Phys. Rev. D}. {\bf 7},
  \penalty0 2333--2346,  (1973).

\bibitem{Kopp:2007}
A.~Kopp, X.~Jia, and S.~Chakravarty, Replacing energy by von {N}eumann entropy
  in quantum phase transitions, \emph{Ann. Phys.} {\bf 322}, \penalty0
  1466--1476,  (2007).

\bibitem{Edwards:1971}
J.~T. Edwards and D.~J. Thouless, Regularity of density of states in
  {A}nderson{'}s localized electron model, \emph{J. Phys. \textrm{C}}. {\bf 4},
  \penalty0 453,  (1971).

\bibitem{Amico:2008}
L.~Amico, R.~Fazio, A.~Osterloh, and V.~Vedral, Entanglement in many-body
  systems, \emph{Rev. Mod. Phys.} {\bf 80}, \penalty0 517,  (2008).

\bibitem{Zanardi:2002}
P.~Zanardi, Quantum entanglement in fermionic lattices, \emph{Phys. Rev.
  \textrm{A}}. {\bf 65}, \penalty0 042101,  (2002).

\bibitem{Wegner:1980}
F.~Wegner, Inverse participation ratio in $(2 + \epsilon)$-dimensions, \emph{Z.
  Phys. \textrm{B}}. {\bf 36}, \penalty0 209--214,  (1980).

\bibitem{Laughlin:1999}
R.~B. Laughlin, Nobel lecture: Fractional quantization, \emph{Rev. Mod. Phys.}
  {\bf 71}, \penalty0 863,  (1999).

\bibitem{Castellani:1986}
C.~Castellani and L.~Peliti, Multifractal wavefunction at the localisation
  threshold, \emph{J. Phys. A}. {\bf 19}, \penalty0 L429--L432,  (1986).

\bibitem{Evers:2008}
F.~Evers and A.~D. Mirlin, Anderson transitions, \emph{Rev. Mod. Phys.} {\bf
  80}, \penalty0 1355,  (2008).

\bibitem{Vasquez:2008}
L.~J. Vasquez, A.~Rodriguez, and R.~A. R\"omer, Multifractal analysis of the
  metal-insulator transition in the three-dimensional anderson model. {I}.
  {S}ymmetry relation under typical averaging, \emph{Phys. Rev. B}. {\bf 78},
  \penalty0 195106,  (2008).

\bibitem{Jia:2008}
X.~Jia, A.~R. Subramaniam, I.~A. Gruzberg, and S.~Chakravarty, Entanglement
  entropy and multifractality at localization transitions, \emph{Phys. Rev. B}.
  {\bf 77}, \penalty0 014208,  (2008).

\bibitem{Slevin:2001}
K.~Slevin, P.~Markos, and T.~Ohtsuki, Reconciling conductance fluctuations and
  the scaling theory of localization, \emph{Phys. Rev. Lett.} {\bf 86},
  \penalty0 3594--3597,  (2001).

\bibitem{Goswami:2007}
P.~Goswami, X.~Jia, and S.~Chakravarty, Quantum {H}all plateau transition in
  the lowest landau level of disordered graphene, \emph{Phys. Rev. B}. {\bf
  76}, \penalty0 205408,  (2007).

\bibitem{Kramer:1996}
B.~Kramer and M.~Schreiber.
\newblock Transfer-matrix methods and finite-size scaling for disordered
  systems.
\newblock In eds. K.~H. Hoffmann and M.~Schreiber, \emph{Computational
  Physics}, p. 166. Springer, Berlin,  (1996).

\bibitem{Huckestein:1995}
B.~Huckestein, Scaling theory of the integer quantum {H}all effect, \emph{Rev.
  Mod. Phys.} {\bf 67}, \penalty0 357--396,  (1995).

\bibitem{Ando:1974}
T.~Ando and Y.~Uemura, Theory of quantum transport in a two-dimensional
  electron system under magnetic fields. {I}. {C}haracteristics of level
  broadening and transport under strong fields, \emph{J. Phys. Soc. Jpn.} {\bf
  36}, \penalty0 959,  (1974).

\bibitem{Pfeuty:1970}
P.~Pfeuty, The one-dimensional ising model with a transverse field, \emph{Ann.
  Phys. ({NY})}. {\bf 57}, \penalty0 79--90,  (1970).

\bibitem{Osborne:2002}
T.~J. Osborne and M.~A. Nielsen, Entanglement in a simple quantum phase
  transition, \emph{Phys. Rev. \textrm{A}}. {\bf 66}, \penalty0 032110,
  (2002).

\bibitem{Zurek:1993}
W.~H. Zurek, S.~Habib, and J.~P. Paz, Coherent states via decoherence,
  \emph{Phys. Rev. Lett.} {\bf 70}, \penalty0 1187--1190,  (1993).

\bibitem{Refael:2009}
G.~Refael and J.~E. Moore, Criticality and entanglement in random quantum
  systems, \emph{J. Phys. A}. {\bf 42}, \penalty0 504010,  (2009).

\bibitem{Chakravarty:1999}
S.~Chakravarty, S.~Kivelson, C.~Nayak, and K.~Voelker, Wigner glass, spin
  liquids and the metal-insulator transition, \emph{Phil. Mag. B}. {\bf 79},
  \penalty0 859--868,  (1999).

\bibitem{Voelker:2001}
K.~Voelker and S.~Chakravarty, Multiparticle ring exchange in the {W}igner
  glass and its possible relevance to strongly interacting two-dimensional
  electron systems in the presence of disorder, \emph{Phys. Rev. B}. {\bf 64},
  \penalty0 235125,  (2001).

\end{thebibliography}
\end{document}